\documentclass[twocolumn,showpacs,preprintnumbers,amsmath,amssymb]{revtex4}
\usepackage{times}
\usepackage{helvet}
\usepackage{courier}
\usepackage{graphicx}

\long\def\comment#1{}

\newcommand{\remove}[1]{}

\newcommand{\voteTotal}{N_{\mbox{\scriptsize vote}}}
\newcommand{\visitRate}{\nu} 
\newcommand{\frontRate}{\nu_{\mbox{\scriptsize f}}} 
\newcommand{\newRate}{\nu_{\mbox{\scriptsize u}}} 
\newcommand{\friendsRate}{\nu_{\mbox{\scriptsize friends}}} 
\newcommand{\fractionToPage}{f_{\mbox{\scriptsize page}}}
\newcommand{\frontPageGrowth}{k_{\mbox{\scriptsize f}}} 
\newcommand{\newPageGrowth}{k_{\mbox{\scriptsize u}}} 

\newcommand{\erfc}{\mbox{erfc}} 
\newcommand{\rms}{{\sc rms}} 

\newcommand{\hour}{\mbox{hr}} 
\newcommand{\minute}{\mbox{min}}

\begin{document}
\title{Using a Model of Social Dynamics to Predict \\Popularity of News}

\author{Kristina Lerman}
 \email{lerman@isi.edu}
\affiliation{%
USC Information Sciences Institute\\
4676 Admiralty Way, Marina del Rey, CA 90292
}

\author{Tad Hogg}
       \email{tadhogg@yahoo.com}
      \affiliation{HP Labs \\
      1501 Page Mill Road,      Palo Alto, CA 94304, USA
}

\begin{abstract}
Popularity of content in social media is unequally distributed, with some items receiving a disproportionate share of attention from users. Predicting which newly-submitted items will become popular is critically important for both companies that host social media sites and their users. Accurate and timely prediction would enable the companies to maximize revenue through differential pricing for access to content or ad placement. Prediction would also give consumers an important tool for filtering the ever-growing amount of content. Predicting popularity of content in social media, however, is challenging due to the complex interactions among content quality, how the social media site chooses to highlight content, and influence among users. While these factors make it difficult to predict popularity \emph{a priori}, we show that stochastic models of user behavior on these sites allows predicting popularity based on early user reactions to new content. By incorporating aspects of the web site design, such models improve on predictions based on simply extrapolating from the early votes. We validate this claim on the social news portal Digg using a previously-developed model of social voting based on the Digg user interface.
\end{abstract}

\maketitle

%
%

\section{Introduction}
Success or popularity in social media is not evenly distributed. Instead, a small number of users dominate the activity on the site, and receive most of the attention of other users. The popularity of contributed items also shows this extreme diversity.
Relatively few of the four billion images on the social photo-sharing site Flickr, for example, are viewed thousands of times, while most of the rest are rarely viewed. Of the more than 16,000 new stories submitted to the social news portal Digg every day, only a handful go on to become wildly popular, gathering thousands of votes, while most of the remaining stories never receive more than a single vote from the submitter herself. Among thousands of new blog posts every day, only a handful rise above the noise.
It is critically important to provide users with tools to help them sift through the vast stream of new content to identify interesting items in a timely manner, or least those items that will prove to be successful or popular.
Accurate and timely prediction will also enable social media companies that host user-generated content to maximize revenue through differential pricing for access to content or ad placement, and encourage greater user loyalty by helping their users quickly find interesting new content.

Success in social media is difficult to predict. Although early and late popularity, which can be measured in terms of the number of views or votes an item generates, are somewhat correlated~\cite{Gomez08,Szabo09}, we know little about what drives success. Is it item's inherent quality~\cite{Agarwal08}, consumer response to it~\cite{CraneSornette08}, or some external factors, such as social influence~\cite{Lerman07digg,Lerman07flickr,Lerman08wosn}? In a landmark study, Salganik et al.~\cite{Salganik06} addressed this question experimentally by measuring the impact of content quality and social influence on the eventual popularity or success of cultural artifacts. They showed that while quality contributes only weakly to their eventual success, social influence, or knowing about the choices of other people, is responsible for both the inequality and unpredictability of success.
In their experiment, Salganik et al. asked users to rate songs they listened to. The users were assigned to different groups. In the control group (independent condition), users were simply presented with lists of songs. In the other group  (social influence condition), users were also shown how many times each song was downloaded by other users. The social influence condition resulted in large inequality in popularity of songs, as measured by the number of times the songs were downloaded.
Although a song's quality, as measured by its popularity in the control group, was positively related to its eventual popularity in the social condition group, the variance in popularity at a given quality was very high, meaning that two songs of similar quality ended up with very different levels of success. Moreover, when users were aware of the choices made by others, popularity was also very unpredictable\remove{ under the social influence condition}.


Although Salganik et al.'s study was limited to a small set of songs created by unknown bands, its conclusions about inequality and unpredictability of success appear to apply to cultural artifacts in general and social media production in particular. While this may at first sound discouraging, as we will show in this paper, a model of social dynamics that includes social influence can help make success in social media predictable. Specifically, we claim that \emph{modeling the collective behavior of users of a social media site allows us to predict the popularity of  items from the users' early reaction to them}. We investigate the claim empirically using data from the social news portal Digg. Digg allows users to submit and collectively moderate news stories by voting on them. Digg selects a hundred or so stories from the thousands that are submitted daily, to feature on its front page. The proprietary promotion algorithm is Digg's way of making a prediction about which stories are interesting to the community and will accumulate many votes.
In previous works, we used the stochastic modeling framework~\cite{Lerman05sab} to mathematically describe social dynamics of Digg users~\cite{Lerman07ic,Hogg09icwsm}. The model, which took into account the user interface and how it affects user behavior, described how the number of votes received by stories changed in time.
We showed qualitative agreement between the data and the model, indicating that the features of the Digg user interface we considered can explain the patterns of collective voting. In this paper we use the model to predict whether a newly submitted story will be promoted based on Digg users' early reaction to it. Moreover, we use the model to predict how popular or successful the story will become, i.e., how many votes it will receive.
The stochastic modeling framework is general and can be applied to other social media sites, making prediction of popularity of content on those sites possible.

The paper is organized as follows. In Section~\ref{sec:digg} we describe details of Digg.  In Section~\ref{sec:model} we summarize the model developed in earlier works. Next, in Section~\ref{sec:prediction} we show how  this model can predict eventual popularity of newly submitted stories on Digg. We discuss results  in Section~\ref{sec:discussion} and compare against other prediction methods outlined in Section~\ref{sec:related}.

\section{Social news portal Digg}
\label{sec:digg}
With over 3 million registered users, the social news aggregator Digg is one of the more popular news portals on the Web.  Digg allows users to submit and rate news stories by voting on, or `digging', them. There are many new submissions every minute, over 16,000 a day. Every day Digg picks about a hundred stories that it deems to be \emph{popular} and promotes them to the front page. Although the exact promotion mechanism is kept secret and changes occasionally, it appears to take into account the number of votes the story receives and how rapidly it receives them. Digg's success is fueled in large part by the emergent front page, which is created by the collective decision of its many users.

\begin{figure}[tbh]
      \includegraphics[width=3.3in]{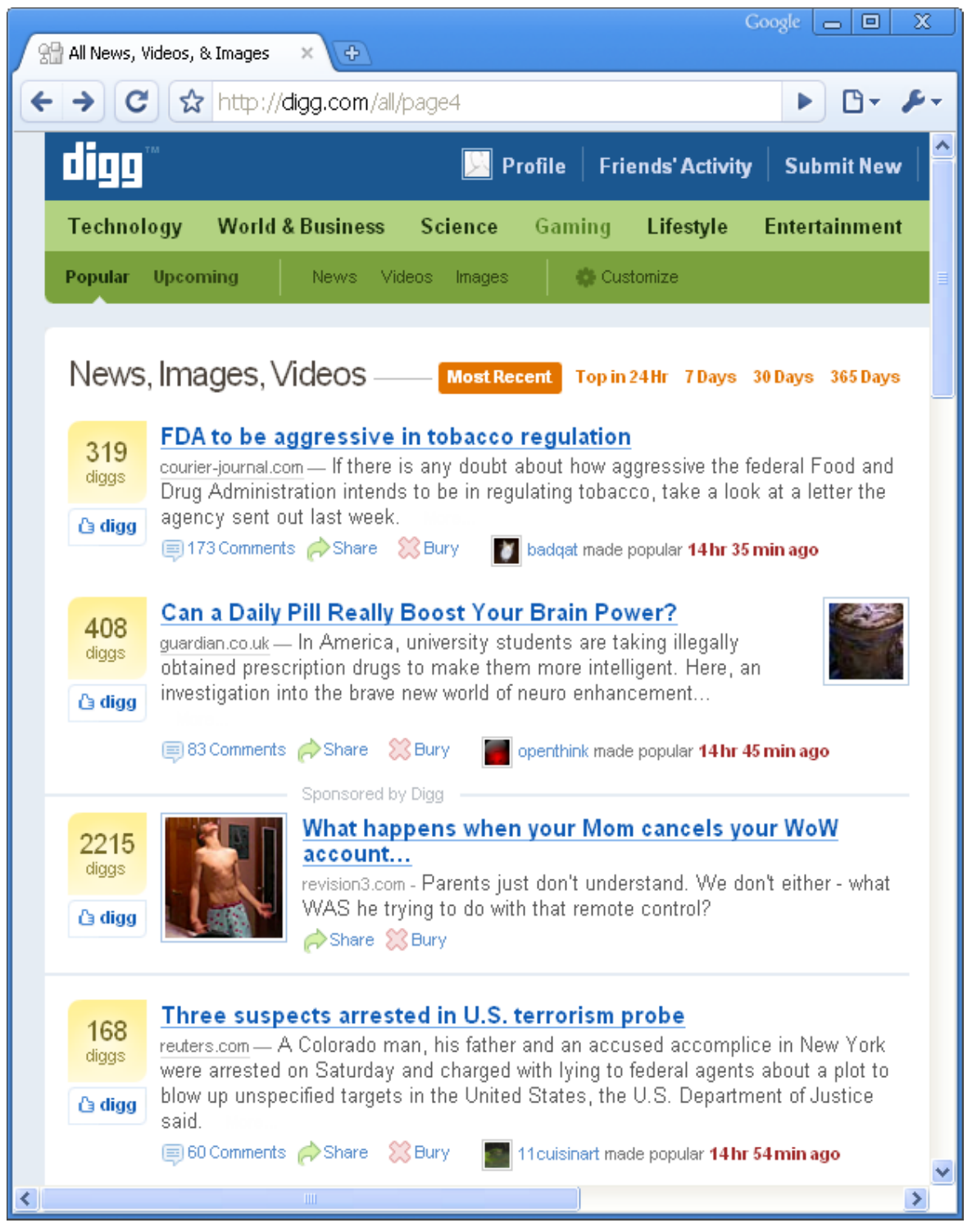} \\
  \caption{Screenshot of the front page of the social news aggregator Digg.}\label{fig:screenshot}
\end{figure}

\subsection{User interface}
A newly submitted story goes to the \emph{upcoming} stories list, where it remains for 24 hours, or until it is promoted to the front page, whichever comes first. Newly submitted stories are displayed as a chronologically ordered list, with the most recently submitted story at the top of the list, 15 stories to a page. To see older stories, a user must navigate to the upcoming stories page 2, 3, etc. Promoted stories (Digg calls them `popular') are also displayed as a list on the front pages, 15 stories to a page, with the most recently promoted story at the top of the list. To see older stories, user must navigate to front page 2, 3, etc. Figure~\ref{fig:screenshot} shows a screenshot of a Digg front page.

Digg also allows users to designate friends and track their activities, i.e., see the stories friends recently submitted or voted for. The \emph{friends interface} is available through the Friends Activity link at the top of any Digg web page (see, for example, Fig.~\ref{fig:screenshot}). The friend relationship is asymmetric. When user $A$ lists user $B$ as a \emph{friend}, $A$ can watch the activities of $B$ but not vice versa. We call $A$ the \emph{fan} of $B$. A newly submitted story is visible in the upcoming stories list, as well as to submitter's fans through the friends interface. With each vote, a story becomes visible to the voter's fans through the friends interface, which shows the newly submitted stories that user's friends voted for.

In addition to these interfaces, Digg also allows users to view the most popular stories from the previous day, week, month, or a year. Digg also implements a social filtering feature which recommends stories, including upcoming stories, that were liked by users with a similar voting history. This interface, however, was not available at the time the data for our study was collected.

\begin{figure}[tbh]
  \begin{tabular}{c}
      \includegraphics[width=3.0in]{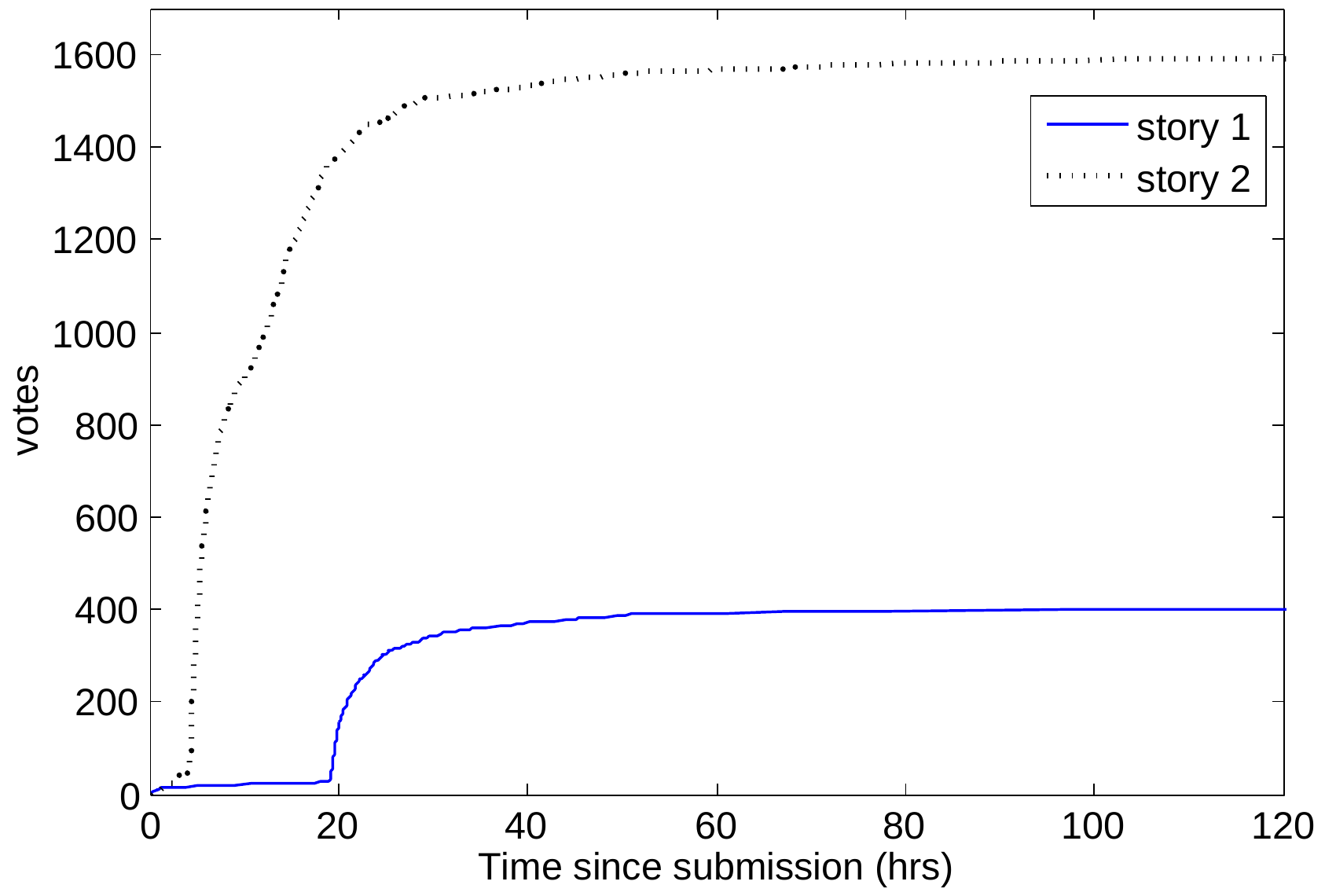} \\
      (a) \\
      \includegraphics[width=3.2in]{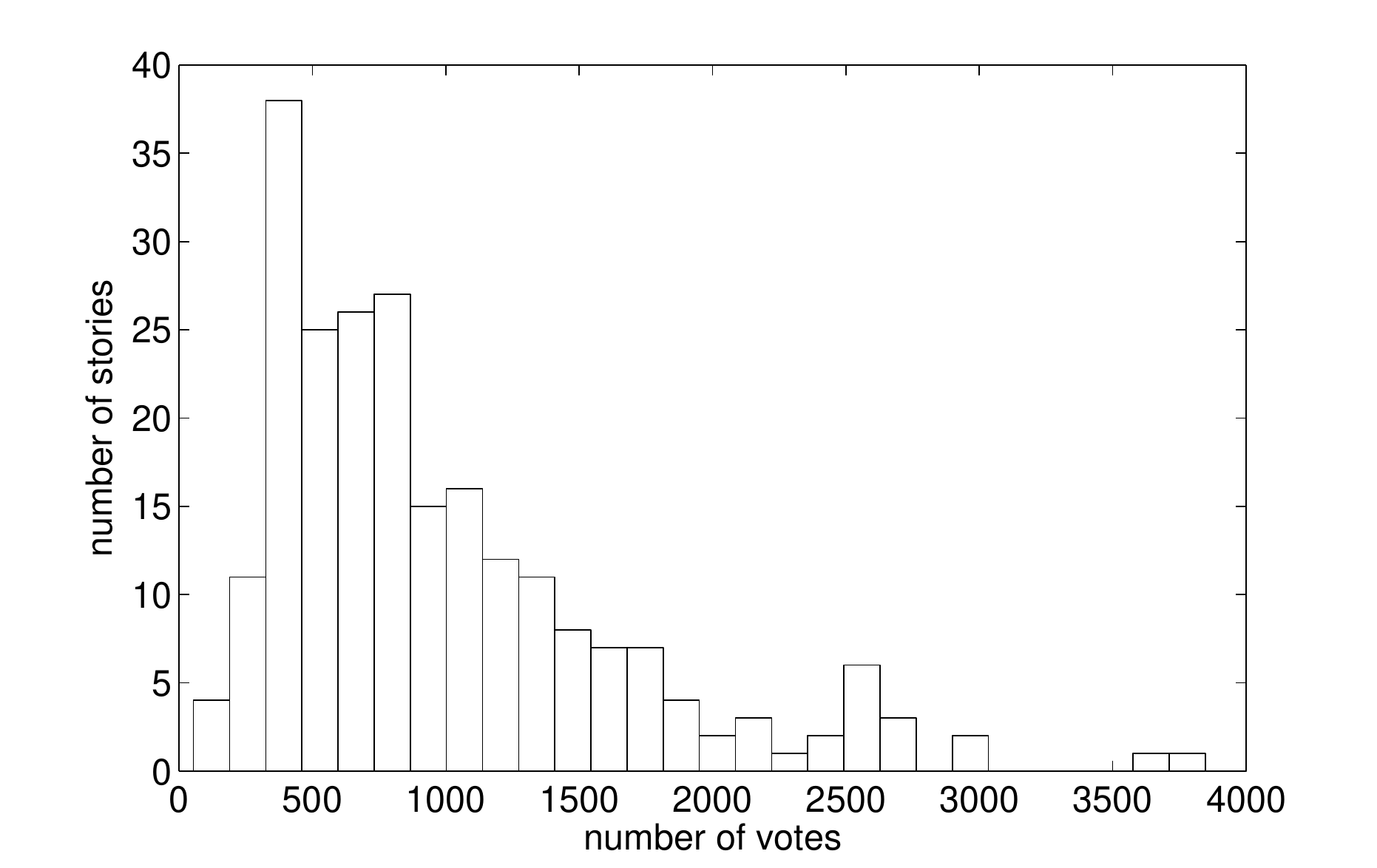}\\
        (b)
  \end{tabular}
  \caption{Dynamics of social voting. (a) Evolution of the number of votes received by two front page stories in June 2006. (b) Distribution of popularity of 201 front page stories submitted in June 2006.}\label{fig:votes}
\end{figure}

\subsection{Inequality of popularity}
While a story is in the upcoming stories list, it accrues votes slowly. After it is promoted to the front page, it accumulates votes at a much faster pace. For example,  Fig.~\ref{fig:votes}(a) shows the evolution of the number of votes for two stories submitted in June 2006. The point where the slope abruptly increases corresponds to promotion to the front page. As the story ages, accumulation of new votes slows down~\cite{Wu07}, and after a few days the total number of votes received by a story saturates to some value. This value,\remove{The total number of votes the story receives by that time} which we also call the final number of votes, gives a measure of the story's success or {popularity}.

Popularity varies widely from story to story. Figure~\ref{fig:votes}(b) shows the distribution of the final number of votes received by front page stories that were submitted over a period of about two days in June 2006. The distribution is characteristic of `inequality of popularity', since a handful of stories become very popular, accumulating thousands of votes, while most others can only muster a few hundred votes. This distribution applies to front page stories only. Stories that are never promoted to the front page receive very few votes, in many cases just a single vote from the submitter.

While the exact shape of the distribution differs among social media sites, the long tail is  a  ubiquitous feature~\cite{Anderson06} of {human activity} \remove{social content production}. It is present in inequality of popularity of cultural artifacts, such as books and music albums~\cite{Salganik06}, and also manifests itself in a variety of online behaviors, including tagging, where a few documents are tagged much more frequently than others,  collaborative editing on wikis~\cite{Kittur06}, and general social media usage~\cite{Wilkinson08}.
\remove{
}
While unpredictability of popularity is more difficult to verify than in the controlled experiments of Salganik et al., it is reasonable to assume that a similar set of stories submitted to Digg on another day will end with radically different numbers of votes. In other words, while the distribution of the  final number of votes  these stories receive will look similar to the distribution in Figure~\ref{fig:votes}(b), the number of votes received by individual stories will be very different in the two realizations.

\subsection{Predictability of popularity}
These observations make predicting popularity of social media content difficult. We claim, however, that we can leverage social influence, the very factor responsible for inequality and unpredictability of popularity, to predict the popularity of social media content.
\remove{In Salganik et al.'s experiments,} Social influence occurs when information about the choices or opinions of others affects a user's behavior. In Salganik et al.'s social influence was exerted by showing to a user the number of times a particular item was downloaded. This information affected what items users chose to download, ultimately leading to a large disparity in the number of downloads of specific items. On Digg, social influence manifests itself through the friends interface, which shows users the stories their friends chose to vote for. In previous works~\cite{Lerman07ic,Hogg09icwsm} we have constructed a mathematical model of the dynamics of social voting on Digg that takes social influence into account. We showed that the model explains the evolution of the number of votes received by Digg stories. In this paper we use the model to predict the popularity of  newly submitted stories. Specifically, we use the model to estimate the inherent quality of a new story from the Digg users' early reaction to it. Next, using this estimate, we predict the story's final number of  votes. In the sections below we summarize the model and validate it on a sample of stories retrieved from Digg.

\section{Social dynamics of Digg}
\label{sec:model}
The model of the dynamics of social voting on Digg~\cite{Lerman07ic,Hogg09icwsm} is based on the stochastic processes framework~\cite{Lerman05sab}, which represents each Digg user as a stochastic process with a small number of states. For users of a social media site, the states correspond to actions such as \emph{register} for the site, \emph{follow link} to a story, \emph{vote} on the story, \emph{befriend} another user, and so on. This abstraction captures much of the inherent individual complexity by casting individual's decisions as inducing probabilistic transitions between states. The framework allows us to relate aggregate behavior of a group of users, such as voting, to simple descriptions of their individual behavior. In past work, we used the model of social voting to study how individual stories accumulate votes on Digg. In this paper, we use the model to explain \remove{the inequality of the final number of votes received by stories,} why some stories accumulate many more votes than others. In addition to the model's explanatory power, we investigate its predictive power.
We first describe the data sets we collected for our study and then present an overview of the model developed in \cite{Hogg09icwsm}.

\subsection{Data sets}
%
%

We collected data by scraping web pages in Digg's Technology section in May and June 2006. The May data set consists of stories that were submitted to Digg May 25-27, 2006. We followed stories by periodically scraping Digg  to determine the number of votes stories  received as a function of the time since their submission. We collected at least 4 such observations for each of 2152 stories, submitted by 1212 distinct users. Of these stories, 510, by 239 distinct users, were promoted to the front page. We followed the promoted stories over a period of several days.

The June data set consists of 201 stories promoted to the front page between June 27 -- 30, 2006. For each story, we collected the names of the first 216 users who voted on the story. In addition, we also collected information about stories that were submitted to Digg between June 30, 2006 and July 1, 2006. From this set, we retained stories that received at least 10 votes, resulting in 159 stories. In October 2009, we updated information about the front page and upcoming stories, using the Digg API to obtain time stamps of the first (up to 216) votes for each story, the total number of votes it received, and for the stories in the upcoming sample, their promotion time, if it exists.

In addition to data about stories, we also extracted a snapshot of the social network of the top-ranked 1020 Digg users (as of June, 2006). This data contained the names of each user's friends and fans. As a reminder, user $A$'s friends are all the users that $A$ is watching (outgoing links on the social network graph), while $A$'s fans  are all the users watching his activity (incoming links).  Since the original network did not contain information about all the voters in the June data set, we augmented it in February 2008 by extracting names of friends of more than $15,000$ additional users.
Many of these users added friends between June 2006 and February 2008. Although Digg does not provide information about the time the new link was created on its web page, it does list these links in reverse chronological order, with the most recent link appearing on top. In addition to friend's name, Digg also gives the date friend joined Digg. By eliminating friends who joined Digg after June 30, 2006, we believe we were able to faithfully reconstruct the fan links for all voters in our data set.
Note that the fans network in the two data sets was slightly different. In the May data set, we retained the number of fans for the top 1020 users, and assumed that other users had zero fans. In the June data set, we know who active users (who voted recently) list as friends and calculate the number of active fans for each submitter. Both are reasonable interpretations of the number of fans, and the exact meaning of the number of fans should depend on the application.

\subsection{Dynamical model of social voting}
When a user visits Digg, she can choose to browse its \emph{front}
pages to see recently promoted stories, \emph{upcoming} stories
pages to see recently submitted stories, or use the \emph{friends}
interface to see the stories her friends have recently submitted or
voted for. She can select one of the stories to read, and depending
on whether she considers it interesting, \emph{vote} for it.
Alternatively, after perusing Digg's pages, she may choose to leave
it. The user's environment, the stories she is seeing, is
itself changing in time depending on actions of all other users.

At an aggregate level, we focus on how the number of votes a story
receives changes over time. The changing state of a story is
characterized by three values: the number of votes, $\voteTotal(t)$,
the story has received by time $t$ after it was submitted to Digg,
the list the story is in at time $t$ (upcoming or front pages) and
its location within that list, which we denote by $q$ and $p$ for
upcoming and front page lists, respectively.

Stochastic modeling provides a framework for relating users' individual choices \remove{, specified by the diagram in Fig.~\ref{fig:user-fsa},} to  their aggregate behavior, which is, in turn, related to the changes in the state of a single story. The aggregate user behavior on Digg at a given time has the following components: the number of
users who see a story via one of the front pages, one of the
upcoming pages, through the friends pages, and number of users who
vote for a story, $\voteTotal$.
In other words, the votes a story receives depends on the combination of its visibility and interest, with visibility coming from different parts of the Digg user interface: the friends interface, upcoming and front page lists, and the position within each list.
The Rate Equation for $\voteTotal(t)$ is:
\begin{equation}\label{eq:diggs}
    \frac{d \voteTotal(t)}{d t} =r ( \frontRate(t) + \newRate(t) + \friendsRate(t) )
\end{equation}
\noindent where $r$ measures how \emph{interesting} the story is, i.e., the
probability a user seeing the story will vote on it, and
$\frontRate$, $\newRate$ and $\friendsRate$ are the rates at which
users find the story via one of the front or upcoming pages, and
through the friends interface, respectively.

\remove{
We can directly measure some aspects of the effect the user interface has on story evolution by studying a characteristic sample of stories. These measurements provide the parameters necessary to link the model to the dynamics of votes received by actual stories. Story interest to users, specified by parameter $r$, depends on inherent characteristics of a story, which we cannot now directly measure. Instead, we leave $r$ as a parameter in the model and estimate it based on the observed evolution of votes.
}

To solve Eq.~\ref{eq:diggs}, we must model the rates at
which users find the story through the different parts of the Digg interface.
These rates depend on the story's location in each list (upcoming or front page) and how users navigate to that position in the list. While many
details of these behaviors are not readily observable, we are able to estimate the values required for our model from the sample of data obtained from Digg and by making some reasonable assumptions.  For example, while we do not know how many users visit Digg each day, we assume that a Digg visitor sees the front page first.  The upcoming stories list is less popular than the front page. We model this by assuming that a fraction $c<1$ of Digg visitors proceed to the upcoming stories pages.

Story position depends on the details of Digg user interface. Digg splits each story list into groups of 15 stories, with 15 most recently submitted (promoted) stories on the first upcoming (front) page, the next group of 15 on the second page, and so on. We model this process as decreasing visibility as a function of location, the value of $\fractionToPage(p)$, through $p$ taking on fractional
values. Thus, $p=1.5$ denotes the position of a story half way down the first page of the list.
Values of $p$ and $q$ grow linearly in time as new stories are promoted to the front page and submitted to the upcoming stories list.

In addition to story position in the list, we need a description of how users navigate to that position. While we do not have data about Digg visitors' behavior,
specifically, how many proceed to page 2, 3 and so on, generally
when presented with lists over multiple pages on a web site,
successively smaller fractions of users visit later pages in
the list. Following~\cite{huberman98}, we use an inverse Gaussian to model the distribution of
the number of pages a user visits before leaving the web site. We model the decreasing visibility of stories as they move down the list on a given page through $p$ and $q$ taking on fractional values in the inverse Gaussian model of user navigation.
\remove{
Therefore, to model the visibility of a story on page $p$ of the front or upcoming
stories lists, the relevant distribution is the fraction of users who visit
\emph{at least} $p$ pages, i.e., the upper cumulative distribution
of the inverse Gaussian:
\begin{equation}
\fractionToPage(p) = \frac{1}{2}\left( F_p(-\mu) - e^{2\lambda/\mu}
F_p(\mu) \right)
\end{equation}
where $F_p(x)=\erfc(\alpha_p (p-1+x)/\mu)$, $\erfc$ is the
complementary error function, $\alpha_p = \sqrt{\lambda/(2(p-1))}$,
and with $\fractionToPage(1)=1$.
}

When a story is promoted, it becomes visible at the top of the front page list. An accurate model of this process would require us to reverse engineer Digg's promotion algorithm. Instead, we use a simple threshold to model how a story is promoted to the front page. The threshold model appears to approximate Digg's promotion algorithm well, and works as follows.
Initially the story is visible on the upcoming stories pages. When the number of accumulated votes exceeds a
promotion threshold $h$, the story moves to the front page.

Next, we model story's visibility through the friends interface. We only consider two components of the friends interface, which allow users to see stories their friends (\emph{i}) submitted or (\emph{ii}) voted for in the preceding 48 hours. Fans of the story's submitter can find the story via the friends interface at any time after submission, regardless of which list it is on. As additional users vote on the story, their fans can
also see the story through the friends interface, regardless of the list the story is on. We model this with $s(t)$, the number of fans of
voters on the story by time $t$ who have not yet seen the story.
We suppose these users visit Digg daily,
and since they are likely to be geographically distributed across
all time zones, the rate fans discover the story is distributed over
the day. A simple model of this behavior takes fans arriving at the
friends page independently at a rate $\omega$. As fans read the
story, the number of potential voters gets smaller, i.e., $s$
decreases at a rate $\omega s$. At the same time, the number of additional fans who can see the story through the friends interface grows as $\Delta s = a \voteTotal^{-b}$ for each new vote, with $a=51$ and $b=0.62$.
Combining these models of growth in the expected number of available fans and
its decrease as fans return to Digg, we have
\begin{equation}
\label{eq:fans}
\frac{d s}{d t} = -\omega s + a \voteTotal^{-b} \frac{d
\voteTotal}{d t}
\end{equation}
with initial value $s(0)$ equal to the number of fans of the story's
submitter, $S$.

In summary, the rates in Eq.~\ref{eq:diggs} are:
\begin{eqnarray*}
  \frontRate &=&  \visitRate \fractionToPage(p(t)) \, \Theta(\voteTotal(t)-h) \\
  \newRate &=& c \, \visitRate \fractionToPage(q(t)) \, \Theta(h-\voteTotal(t)) \Theta(24\hour-t)\\
  \friendsRate &=& \omega s(t)
\end{eqnarray*}
\noindent where $t$ is time since the story's submission and
$\visitRate$ is the rate users visit Digg. The first step function
in $\frontRate$ and $\newRate$ indicates that when a story has fewer
votes than required for promotion, it is visible in the upcoming
stories pages; and when $\voteTotal(t)>h$, the story is visible on
the front page. The second step function in $\newRate$ accounts for
a story staying in the upcoming queue for at most $24$ hours.

We solve Eq.~\ref{eq:diggs} subject to initial condition
$\voteTotal(\remove{t=}0)=1$, because a newly submitted story
appears on the top of the upcoming stories queue and it starts with
a single vote, from the submitter.

\subsection{Model parameters and solutions}

\begin{table}[t]
\begin{center}
\begin{tabular}{l|l}
\hline parameter & value \\
\hline rate general users come to Digg & $\nu=10\,\mbox{users}/\minute$ \\
fraction viewing upcoming pages  & $c=0.3$ \\
rate a voters' fans come to Digg & $\omega=0.002/\minute$ \\
page view distribution  & $\mu=0.6$, $\lambda=0.6$ \\
fans per new vote & $a=51$, $b=0.62$ \\
vote promotion threshold    & $h=40$ \\
upcoming stories list location  & $\newPageGrowth = 0.06\,\mbox{pages}/\minute$ \\ 
front page list location  & $\frontPageGrowth = 0.003\,\mbox{pages}/\minute$ \\ 
\hline \multicolumn{2}{c}{story specific parameters} \\
interestingness    & $r$ \\
number of submitter's fans  & $S$ \\
\end{tabular}
\end{center}
\caption{Model parameters. Parameters specifying page view distribution are defined in \protect\cite{Hogg09icwsm}.}\label{parameters}
\end{table}

As shown in \cite{Hogg09icwsm} solutions to Eq.~\ref{eq:diggs} agree with the evolution of votes received by actual stories on Digg. The solutions depend on the model parameters, of which only two parameters --- the story's interestingness $r$ and number of fans of the submitter $S$ --- change from story to story.  We estimated $r$ from the data as the value that minimizes the root-mean-square (\rms) difference between the observed votes and the model predictions. The remaining parameters, given in Table~\ref{parameters}, are fixed.
As described in more detail in \cite{Hogg09icwsm}, some of these parameters, such as the growth in list location, promotion threshold and fans per new vote, were measured directly from the May data set. Other parameters were estimated based on model predictions. The small number of stories in our data set, as well as the approximations made in the model, do not give strong constraints on these parameters. We selected values to give a reasonable match to our observations. These parameters could in principle be measured independently from aggregate behavior with more detailed information on user behavior.

\begin{figure}[t]
\begin{center}
\includegraphics[width=3.2in]{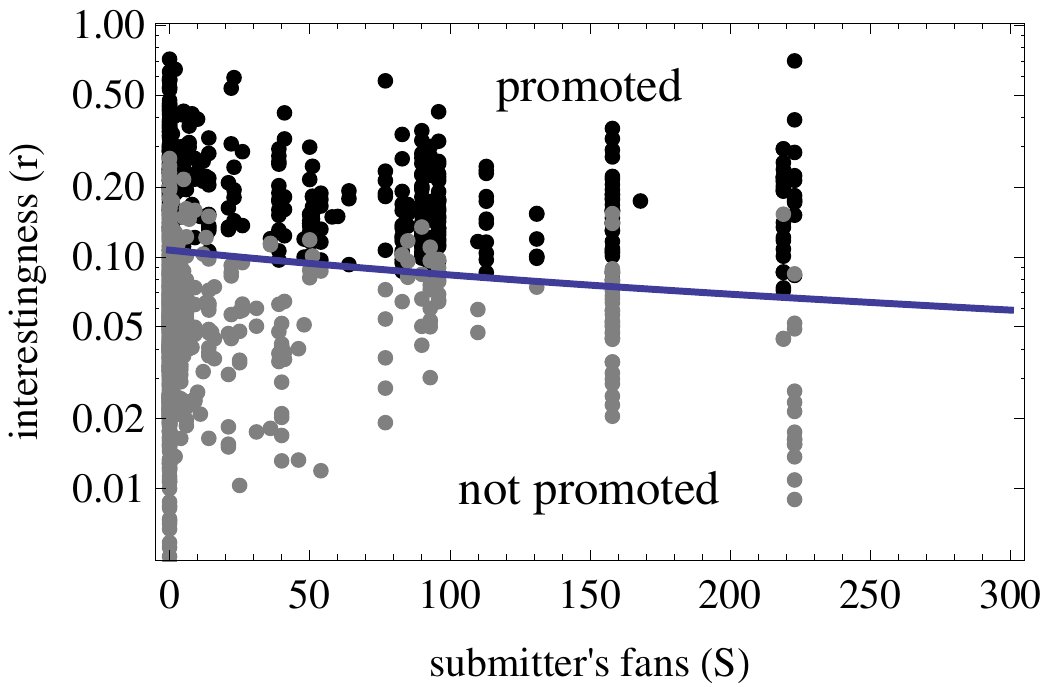}
\end{center}
\caption{Story promotion as a function of $S$ and $r$ for stories in the May data set. The $r$
values are shown on a logarithmic scale. The model predicts stories
above the curve are promoted to the front page. The points show the
$S$ and $r$ values for the stories in our data set: black and gray
for stories promoted or not, respectively.}\label{fig:promotion}
\end{figure}

Fig.~\ref{fig:promotion} shows parameters $r$ and $S$ required for a story to reach the front page according to the model, and how that prediction compares to the stories in the May data set. The model's prediction of whether a story is promoted is correct for $95\%$ of the stories in our data set.
For promoted stories, the correlation between $S$ and $r$ is
$-0.13$, which is significantly different from zero ($p$-value less
than $10^{-4}$ by a randomization test).
Thus a story submitted by a poorly connected user (small $S$) tends
to need high interest (large $r$) to be promoted to the front
page~\cite{Lerman07digg}.

Parameter $r$ depends on the inherent story quality, which we cannot directly measure from our data. However, our interpretation of $r$ as how `interesting' a story is to users appears to be consistent with treating it as representing intrinsic story quality. Specifically, the model reproduces three general observations about behavior of stories on Digg: (1) slow initial growth in votes while the story is on the upcoming list, as shown in Fig.~\ref{fig:votes}(a); (2) more interesting stories (higher $r$) are promoted to the front page faster and receive more votes than less interesting stories; (3)  however, as supported also by observations in \cite{Lerman07digg}, better connected users (high $S$) are more successful in getting their less interesting stories (lower $r$) promoted to the front page than poorly-connected users. These observations give us confidence that the model captures the important details of social voting on Digg.

By estimating $r$ from the observed dynamics of social voting, our model allows us to \emph{separate story quality from social influence and study how each affects the popularity of stories on Digg}.
While there are alternative ways to measure the effects of quality and social influence, they may not be feasible for social media applications. Quality, for example, may be measured through controlled experiments, as in \cite{Salganik06}. Social influence may be measured through surveys or interviews with participants, which is also not usually practical in social media. An empirically grounded model, on the other hand, allows us to quantitatively characterize the effects of quality and social influence on the popularity of social media content, and deduce the strength of these effects from the observed dynamics of popularity. This leads to an insight that models can be used to predict popularity of content.
Specifically,  {observing the initial stages of voting on Digg, and knowing how users are connected, enables us to use the model of social dynamics to estimate $r$, and then use this value to predict how many votes the story will receive in the long-term}.
In the sections below we investigate the implications of the model for determining quality of stories submitted to Digg, and also for predicting the number of votes they will receive. Since the stochastic modeling framework on which the approach is based is general, and has been applied to several other systems~\cite{Lerman05sab,Hogg09}, we conjecture that this approach can also be used to predict popularity of content on other social media sites.

\section{Model-based Prediction}
\label{sec:prediction}

By separating the impact of story quality and social influence on the popularity of stories on Digg, a model of social dynamics enables two novel applications: (1) estimating inherent story quality from the evolution of its observed popularity, and (2) predicting its eventual popularity based on the early reaction of users to the story. We investigate these problems on real-world data extracted from Digg.

\subsection{Estimating story quality}
We can estimate how interesting a story is by comparing the model's solutions to the observed popularity of the story. We take as story interestingness the value of $r$ that minimizes \rms difference between the observed number of votes and the number of votes predicted by the model at the end of the data sample or two days after submission, whichever was earlier.
For the 510 promoted stories in the May data set, the \rms\ relative error between the
number of votes and the model prediction is $14\%$, corresponding to
a \rms\ error of $109$ votes.
For stories not promoted these values are $14\%$ and $1.1$ votes, respectively.

\begin{figure}[t]
\begin{center}
  \begin{tabular}{c}
  \includegraphics[width=3.2in]{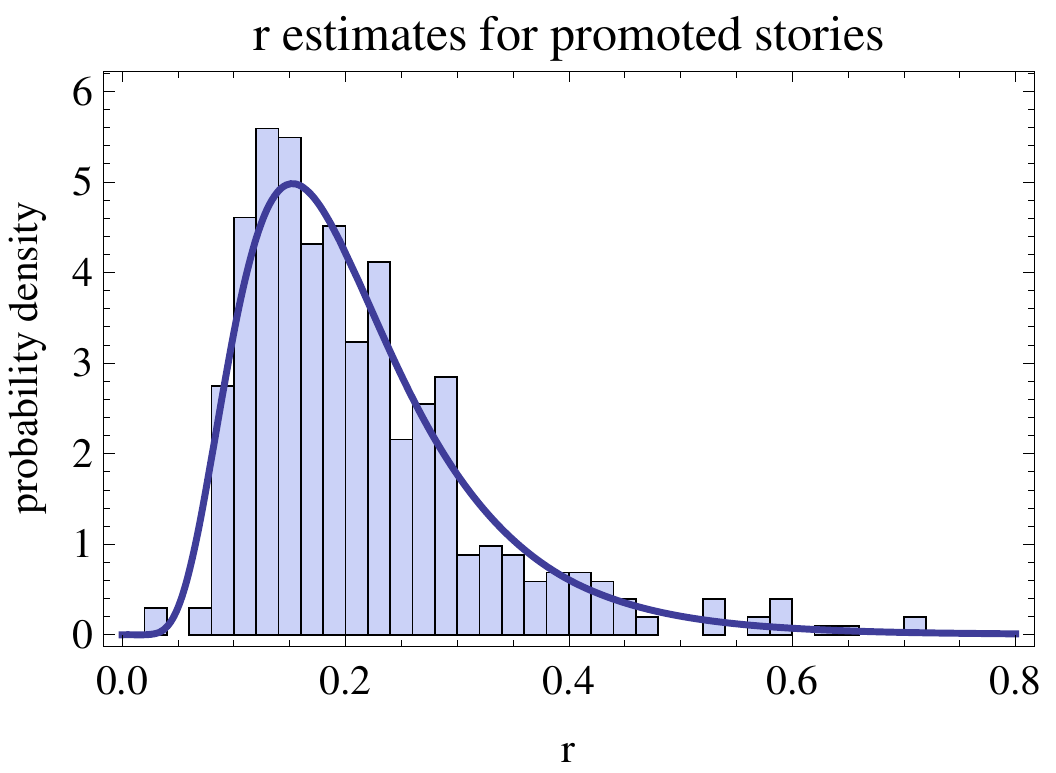} \\
  (a)\\
  \includegraphics[width=3.2in]{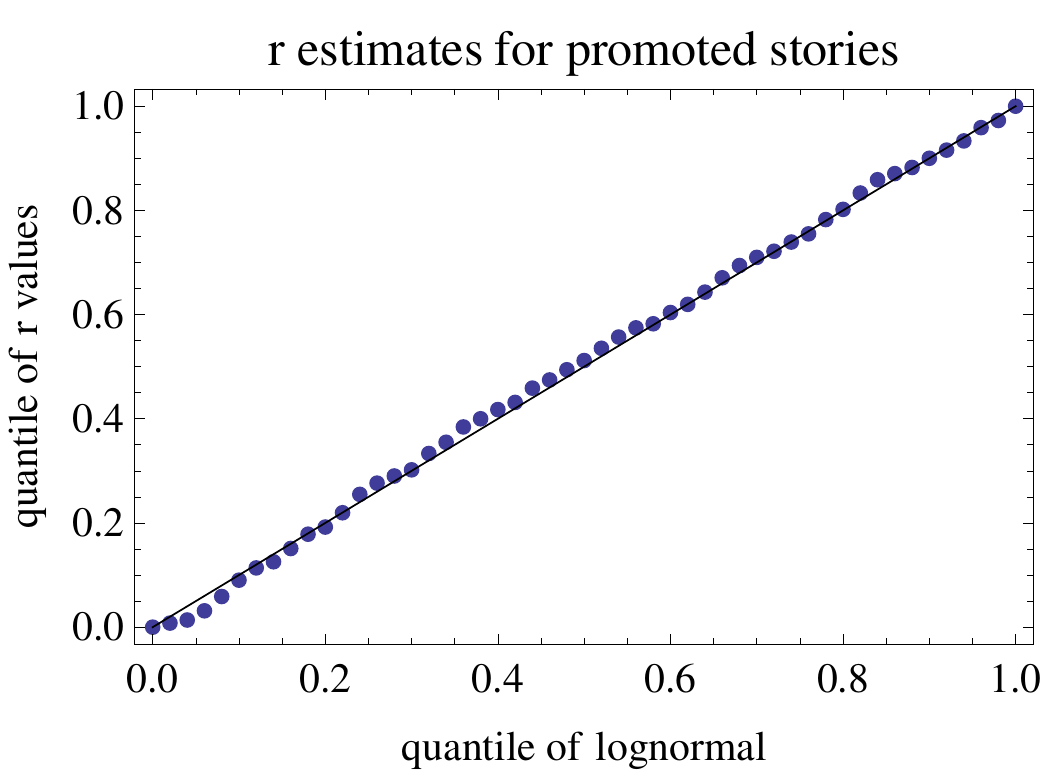} \\
  (b)
  \end{tabular}
\end{center}
\caption{(a) Histogram of estimated $r$ values for the promoted
stories in our data set compared with the best fit lognormal
distribution. (b) Quantile-quantile plot comparing observed
distribution of $r$ values with the lognormal distribution fit.}
\label{fig:r-distribution}
\end{figure}

\begin{table*}[tbh]
  \centering
  \setlength{\tabcolsep}{1pt}
  \begin{tabular}{|c|c|l|}
    \hline
\textbf{final votes} & \textbf{estimated	$r$}  &	\textbf{story title} \\
\hline
3054 &	0.71 &	Lego Aircraft Carrier Complete! \\
3388 &	0.70 &	How to Make a Spider from 5 Crisp Dollar Bills (and Scare Waitresses!) \\
3125 &	0.65 &	Things You Didn't Know About Your Body \\
2981 &	0.63 &	25 Worst Tech products of all time \\
2776 &	0.59 &	The Coolest Solar Eclipse Photo You Will Ever See... \\
2748 &	0.59 &	14 year old kid becomes millionaire through online scamming \\
2701 &	0.58 &	X-Men: Last Stand Post-Credits Scene? \\
2327 &	0.58 &	18 Days of Reckless Computing \\
2690 &	0.58 &	First Photos of MIT's \$100 Laptop \\
1310 &	0.57 &	Nintendo Puts \$250 Price Tag on Wii OFFICIAL \\
2204 &	0.54 &	MacBook vent blocked \\
2413 &	0.54 &	Wii will cost less than \$220 \\
\hline
397	 &  0.09 &	Microsoft: ``OpenDocument is Too Slow'' \\
364 &	0.09 &	AMD aims to take 15\% of notebook market this year \\
278 &	0.09 &	New Intel roadmap reveals Conroe L ``solo'', mobile plans \\
300 &	0.09 &	Interactive display system knows users by touch \\
341 &	0.09 &	A DNA Database For All U.S. Workers? \\
540 &	0.08 &	Computer Viruses Monitored via Dynamic Worldmap \\
258 &	0.08 &	New Sensor Technology Looks at Molecular 'Fingerprint' \\
149 &	0.07 &	Supreme Court won't consider Yahoo case \\
247 &	0.07 &	Lambda Table - A high-res tiled LCD table and interaction device \\
642 &	0.03 &	Interactive dining table \\
1204 &	0.03 &	Websites as graphs: Visualizing the DOM Structure of Websites \\
532 &	0.02 &	MIT Technology Review Launches New Micro-documentary Video Series \\
\hline
 \end{tabular}
\caption{Selection of stories from the May data set with the highest and lowest $r$ values. For each story, we show the final number of votes it received, its estimated $r$ value, and its title.}
\label{tbl:may}
\end{table*}

The estimated $r$ values of stories in the May data set show that the
510 promoted stories have a wide range of interestingness to users. 
As shown in Fig.~\ref{fig:r-distribution}, these $r$ values fit well
to a lognormal distribution with maximum likelihood estimates of the
mean and standard deviation of $\log(r)$ equal to $-1.67\pm0.04$ and
$0.47\pm0.03$, respectively, with the ranges giving the $95\%$
confidence intervals. A randomization test based on the
Kolmogorov-Smirnov statistic and accounting for the fact that the
distribution parameters are determined from the
data~\cite{clauset07} shows the $r$ values are consistent with this
distribution ($p$-value $0.35$). 
Table~\ref{tbl:may} shows some of the stories with the highest, as well as lowest, estimated $r$ values. Stories with higher $r$ values include those bound to pique curiosity, such as ``Lego Aircraft Carrier Complete!'' and lists of the ``worst'' and ``coolest''. Among stories with lower $r$ values are more serious stories about science and technology. Unfortunately, it looks like Digg users do not find such stories interesting.

The $r$ values for June data set have a similar lognormal distribution.
%
While broad distributions occur in
many web sites~\cite{Wilkinson08}, using a model of social dynamics allows us to factor out effects of user interface (various components of story visibility) from the overall distribution of story interestingness.
Thus we can identify variations in the stories' inherent interest to users as measured by their inclination to vote on a story they see. These findings indicate that at least part of the inequality in the distribution of final number of votes received by Digg stories (\emph{cf} Fig.~\ref{fig:votes}(b)) can be attributed to the inequality of their inherent interest to users.

\subsection{Predicting final number of votes}

\begin{figure}[t]
\begin{center}
\includegraphics[width=3.2in]{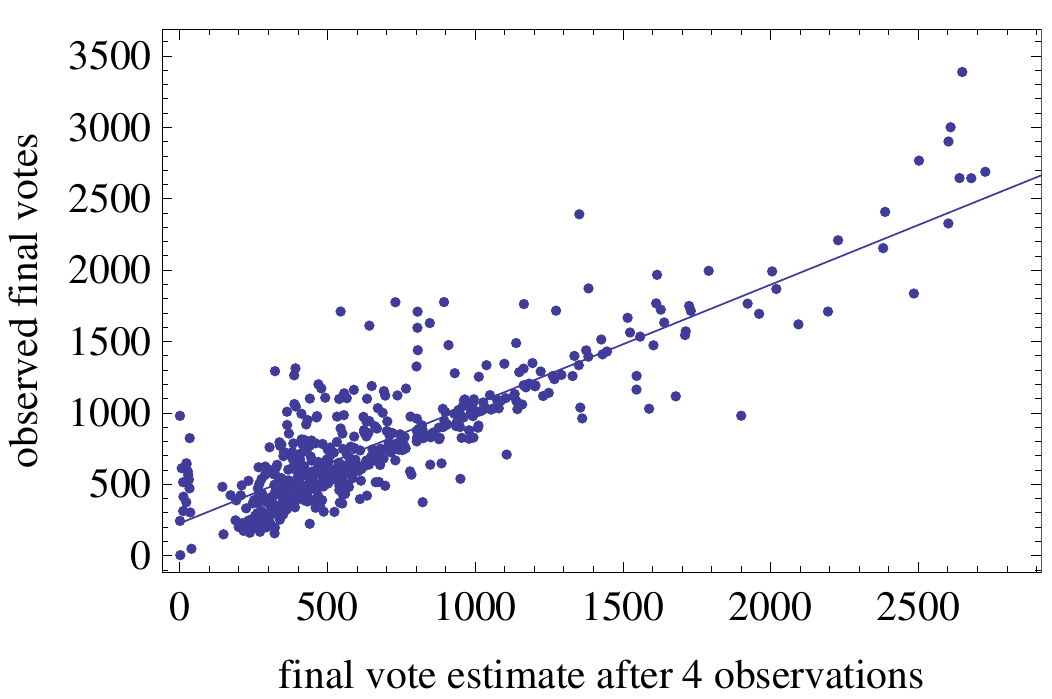}
\end{center}
\caption{Observed number of final votes for promoted stories in the May data set
compared to prediction from the model using the first four
observations of each story to estimate the story's $r$ value. The
line is the best linear fit, with slope $0.84$.}\label{fig:early
estimates}
\end{figure}

Rather than estimating $r$ values from the full voting history, we can estimate them from the early voting history of each story.
For instance, using just the first 4 observations for each
promoted story in the May data set increases the relative error in the votes to $34\%$.
The predicted numbers of votes have $87\%$ correlation with the
observed numbers so early observations provide a strong prediction
of the relative ordering of numbers of votes stories will receive,
as illustrated in Fig.~\ref{fig:early estimates}.
This corresponds to the predictability of eventual ratings from the
early reaction to new content seen on Digg and
You\-Tube~\cite{Szabo09}.

\begin{figure}[t]
\begin{center}
\includegraphics[width=3.2in]{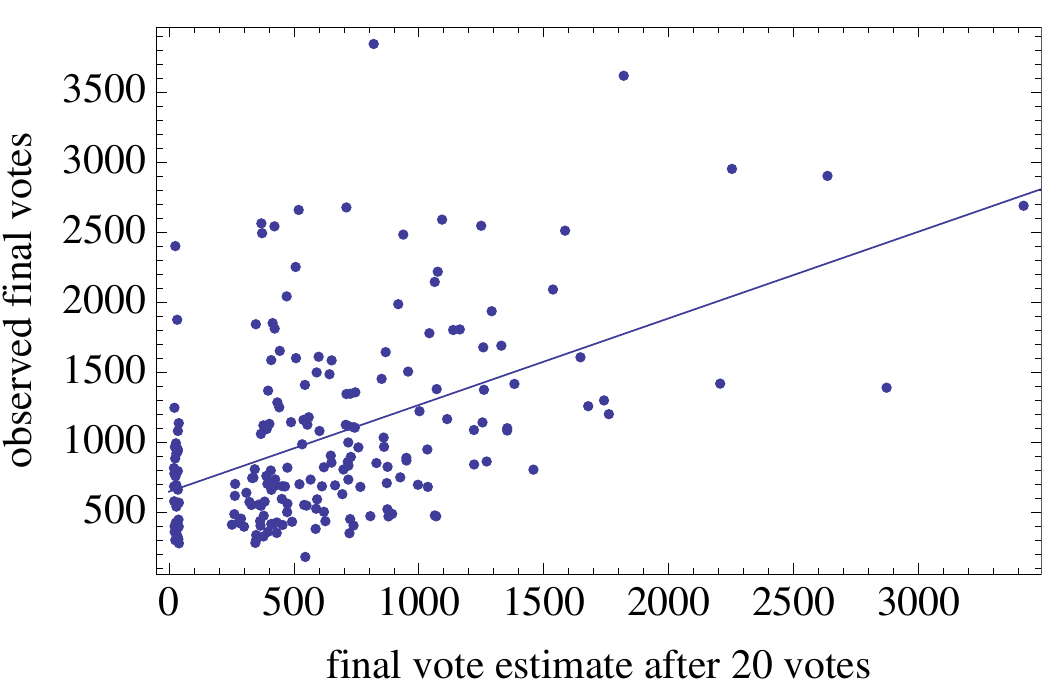}
\end{center}
\caption{Observed number of final votes for promoted stories in the June data set
compared to prediction from the model using the first 20 votes each story received to estimate the story's $r$ value. The
line is the best linear fit, with slope $0.62$.}\label{fig:early
estimates June}
\end{figure}

Figure~\ref{fig:early estimates June} shows predictions for front page stories in the June data set, based on the first 20 votes a story receives and using the model described above, i.e., with parameters determined from the May data set.  In this case, the predictions are not as good (correlation between predicted and actual final votes is $0.49$, the \rms\ error is 593, and the linear fit accounts for only $23\%$ of the variance).

 In both figures, the cluster of points at the extreme left of the plot are promoted stories the model predicts will not be promoted (based on the $r$ estimate from the early votes). Thus their actual final number of votes is considerably larger than the model predicts based on the early votes.


\subsection{Comparing to direct extrapolation}
Once a story reaches the front
page, its subsequent growth in votes is well-predicted from the
number of votes it receives shortly after promotion when accounting
for the hourly and daily variation in story submission
rate~\cite{Szabo09}. However, these predictions apply to promoted stories only and do not take into account changes in visibility of a story through growth in the number of fans.
Although we do not have enough data to reproduce the approach of \cite{Szabo09}, as the first 216 votes often did not cover one hour after promotion required by the approach,
as a simple comparison, we determined the predicted number of
votes based on extrapolating from the rate a story accumulated votes
during the first 4 observations. This simpler model, which does not
consider the number of fans for the story's voters, has a lower
correlation, $75\%$, with the observed numbers and a larger \rms\
error for stories in the May data set. A randomization test comparing these two methods indicates
this reduction in performance is statistically significant
($p$-value less than $5\times 10^{-4}$). Thus,
by incorporating the average growth in number of fans, our model provides a
better description of how stories accumulate votes than simply
extrapolating from early observations while on the upcoming pages.
More generally, by estimating the ``interestingness'' of a story
from early votes, we separate the influence of changing visibility
in the Digg user interface from the underlying rate at which users
will vote on the story if they see it.

Although model-based predictions for stories in the June data set are not as good, nevertheless, using the model improves on direct extrapolation (correlation $0.44$, \rms\ error 610, and fraction of variance $19\%$). We find a similar improvement for predicting the final votes for the upcoming stories of the June data set, e.g., correlation $0.47$ using the model compared to $0.31$ for direct extrapolation.

\begin{figure}[tbh]
\begin{center}
\includegraphics[width=3in]{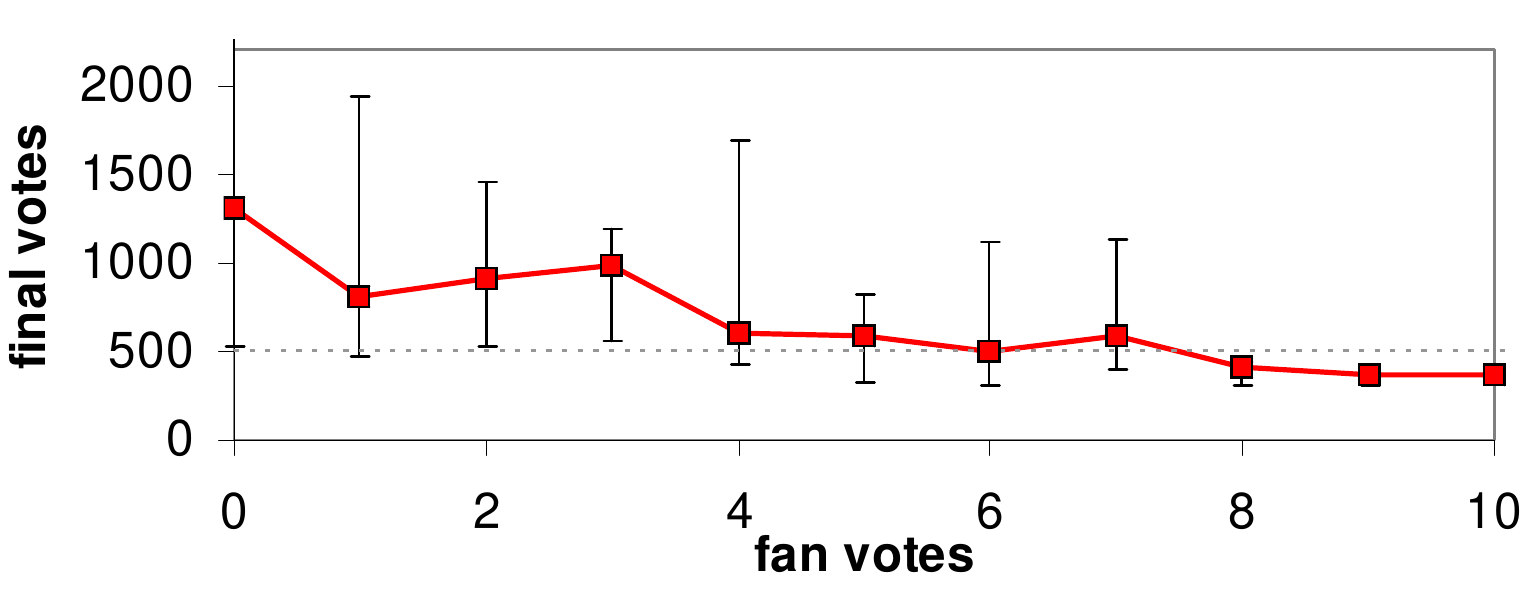}
\end{center}
\caption{Number of fan votes within the first 10 votes vs final votes received by front page stories in the June data set. The dashed line shows 505 votes.}\label{fig:fanvotes}
\end{figure}

\subsection{Comparing to social influence only prediction}
\label{sec:social-influence}
In \cite{Lerman08wosn} we studied the role of social influence in predicting popularity of news stories on Digg. We showed that stories that initially receive many fan votes, i.e., votes from fans of the submitter or previous voters, ultimately go on to accumulate fewer votes than stories that initially receive few fan votes. Although this may at first seem counter intuitive, it is reasonable to expect that a story that is of interest to a narrow community will spread within that community only, while a generally interesting story will spread from many independent sites as users unconnected to previous voters discover it with some small probability and propagate it to their own fans. \cite{Lerman08wosn} did not separate effects of story quality or interestingness from social influence, but simply used the strength of social influence as a predictor of whether the story will receive many votes.

As described in this paper, at the time of submission, a story is only visible on the upcoming stories list and to submitter's fans through the friends interface. As users vote on the story, it becomes visible to their own fans through the friends interface. Some of these fans will find the story interesting and vote for it. Although we cannot confirm it, we assume that if a voter is a fan of the previous voters (including the submitter), social influence, exerted via the friends interface, played a role in helping the voter discover the story.
Therefore, the strength of social influence is measured in terms of the proportion of initial votes that can be made via the friends interface: those coming from the fans of the submitter and previous voters.
Social influence during the early voting period and the final number of votes a story receives are inversely correlated. Figure~\ref{fig:fanvotes} shows the number of fan votes within the first 10 votes vs the final number of votes received by the 201 front page stories in the June data set. The plot shows median number of final votes, with the errors bars showing the distribution of votes, with the outliers removed. Despite wide range of final votes for each value of fan votes, in general, stories that receive relatively few fan votes within the first 10 votes end up becoming very popular, accumulating many hundreds or thousands of votes, while stories that receive many fan votes within the first 10 votes end up with fewer than 500 votes.

We trained a decision tree classifier on front page stories in the June data set to predict whether a story will be successful, i.e., accumulate a large number of votes, based on the strength of social influence during the early stages of voting~\cite{Lerman08wosn}. Each story was characterized by three attributes: number of fan votes it received within the first 10 votes, number of submitter's fans, and a boolean attribute indicating whether the story was successful (i.e., received more than 505 votes).
This classifier can then by used to predict whether a story will become successful by monitoring its spread through the fan network.
As shown in \cite{Lerman08wosn}, the prediction can be made relatively early, after the first 10 votes.

We compare model-based prediction against social influence-based  classifier  described above. We use the classifier to predict whether an upcoming story in the June data set will accumulate more than 505 votes. As argued in \cite{Lerman08wosn}, that prediction should be made for stories submitted by top users, who tend to have bigger and more active fan networks, which make it more difficult for Digg to determine story's general appeal to the rest of its community. There were 39 stories submitted by users who were among  the top-ranked 100 users in June 2006. Of these stories, 13 were actually promoted by Digg, and of these only four went on to receive more than 505 votes. The classifier predicted that 14 of the 39 stories will get more than 505 votes, and of these, only three did. The classifier also predicted that 25 stories will accumulate fewer than 505 votes, and 24 of these predictions were correct. In all, social influence-based classifier correctly predicted the fate of 27 stories. Using the same criterion of success and using only the first 10 votes for prediction, the model-based method predicted that 11 stories will accumulate more than 505, of which 3 did. It also predicted that 28 stories will not reach 505 votes, and 27 of these predictions were correct. In all, model-based method correctly predicted the fate of 30 stories, a $10\%$ improvement over the social influence-based method.

\section{Discussion}
\label{sec:discussion}

There is a number of reasons why predictions for the final number of votes received by June stories were worse than predictions for the stories in the May data set. May data was collected by scraping Digg web pages at regular time interval. While for over half of the promoted and upcoming stories in the May data set the fourth observation was made about four hours since story submission, for many of the remaining stories, 4th observation was made many hours later. Therefore, prediction was able to exploit longer-term dynamics. The first 20 votes used for prediction in the June data set generally accounted for shorter periods since submission. Another reason for the disparity was that the model was calibrated on the May data set. Using parameters calculated from June data could improve predictions. We could not explore this questions due to lack of relevant data. On the other hand, we believe that some prediction accuracy on the June data set demonstrate generalizability of the model. Another difference between the models is that for the May data we used all fans as extracted from Digg, while number of fans in the June data set is based on users who were active (i.e. voted recently). Both definitions seem reasonable to me, so by comparing the May and June results, we're also comparing the use of these different definitions in the two cases.

The model makes several assumptions and approximations which could reduce accuracy of prediction. First, we treated promotion as an exact threshold. Detailed analysis of June data shows this not to be accurate, as some stories were promoted well before they reached 40 votes. The earlier in its history the story is promoted, the more votes it will receive. While we do not know the exact promotion algorithm Digg uses, we can mitigate this problem by giving bounds on the predicted number of votes, which reflect our uncertainty about the promotion mechanism. Another modeling simplification we made is to use growth in the expected number of new fans, given by Eq.~\ref{eq:fans}. Since we know how large the fans network is for each voter, we can compute these values more precisely. This will enable us to treat cases when a vote by a highly connected user, such as \emph{kevinrose}, exposes the story to a large number of users.


Finally, as evidence in Section~\ref{sec:social-influence} suggests, prediction may also benefit from a finer grained model of social influence. While model-based prediction outperforms social influence-only model, we believe that social influence offers valuable evidence about story's interest within and outside a community. Monitoring the spread of interest in a story through the fan network will lead to a better estimate of $r$, which will, in turn, lead to a more accurate prediction of the final number of votes. The value of $r$ could be different to fans vs non-fans. We plan to study these issues in future work.


\section{Related work}
\label{sec:related}

The Social Web provides massive quantities of available data about the behavior of large groups of people. Researchers are using this data to study a variety of topics, including detecting~\cite{Adamic04,Leskovec07kdd} and influencing~\cite{Domingos01,Kempe03} trends in public opinion, and dynamics of information flow in groups~\cite{Wu04,Leskovec07}.

Several researchers examined the role of social dynamics in explaining and predicting distribution of popularity of online content.
Wilkinson~\cite{Wilkinson08} found broad distributions of popularity and user activity on many social media sites and showed that these distributions can arise from simple macroscopic dynamical rules.
Wu and Huberman~\cite{Wu07} constructed a phenomenological model of the dynamics of collective attention on Digg. Their model is parametrized by a single variable that characterizes the rate of decay of interest in a news article. Rather than characterize evolution of votes received by a single story, they show the model describes the distribution of final votes received by promoted stories. Our models offers an alternative explanation for the distribution of votes. Rather than novelty decay, we argue that the distribution can also be explained by the combination of a non-uniform variations in the stories' inherent interest to users and effects of user interface, specifically decay in visibility as the story moves to subsequent front pages. Such a mechanism can also explain the distribution of popularity of photos on Flickr, which would be difficult to characterize by novelty decay.
Crane and Sornette~\cite{CraneSornette08} analyzed a large number of videos posted on You\-Tube and found that collective dynamics was linked to the inherent quality of videos. By looking at how the observed number of votes received by videos changed in time, they could separate high quality videos, whether they were selected by You\-Tube editors or spontaneously became popular, from junk videos. This study is similar in spirit to our own in exploiting the link between observed popularity and content quality. However, while this, and Wu \& Huberman study, aggregated data from tens of thousands of individuals, our method focuses instead on the \emph{microscopic} dynamics, modeling how individual behavior contributes to the observed popularity of content.

Researchers found statistically significant correlation between early and late popularity of content on Slashdot~\cite{Kaltenbrunner07}, Digg and You\-Tube~\cite{Szabo09}. Specifically, similar to our study, Szabo \& Huberman~\cite{Szabo09} predicted long-term popularity of stories on Digg. Through large-scale statistical study of stories promoted to the front page, they were able to predict stories' popularity after 30 days based on their popularity one hour after promotion. Unlike our work, their study did not specify a mechanism for evolution of popularity, and simply exploited the correlation between early and late story popularity to make the prediction. Our work also differs in that we predict popularity of stories shortly after submission, long before they are promoted.
In \cite{Lerman08wosn} we exploited anti-correlation between the number of early fan votes and stories' eventual popularity on Digg. Specifically, we found that stories that initially received few votes from the fans of submitters and previous voters went on to become much more popular than stories which had many initial votes from fans. Using this correlation, we were able to predict whether stories submitted by well connected users would become popular, i.e., receive more than 505 votes. That work exploited social influence only to make the prediction, and the results were not applicable to stories submitted by poorly connected users which were not quickly discovered by highly connected users. In contrast, the approach described in this paper considers effects of social influence regardless of the connectedness of the submitter, and also accounts for story quality in making a prediction about story popularity.

\section{Conclusion}
In the vast stream of new user-generated content, only a few items will prove to be popular, attracting a lion's share of attention, while the rest languish in obscurity. Predicting which items will become popular is exceedingly difficult, even to experts. Research has shown that popularity is weakly related to inherent content quality, and that social influence leads to an uneven distribution of popularity and makes it so difficult to predict. We claim that a model of social dynamics of users on a social media site allows us to quantitatively characterize evolution of popularity of items on that site and study how it is affected by item quality and social influence. We evaluate this claim by studying the social news aggregator Digg, which allows users to submit and vote on news stories. The number of votes a story accumulates on Digg shows its popularity. In an earlier work we developed a model of social voting on Digg, which describes how the number of votes received by a story changes in time. Knowing how interesting a story is and how connected the submitter is fully determines the evolution of the number of votes the story receives. This leads to an insight that a model can be used to predict story's popularity from the initial reaction of users to it. Specifically, we use observations of evolution of the number of votes received by a story shortly after submission to estimate how interesting it is, and then use the model to predict how many votes the story will get after a period of a few days. Model-based prediction outperforms other methods that exploit social influence only, or correlation between early and late votes received by stories. However, results show that we can improve prediction by developing a more fine-grained model that differentiates between how interesting a story is to fans and to the general population.

\section*{Acknowledgments}
We would like to thank Fetch Technologies for providing the tool to extract data from Web pages. In addition we would like to thank Suradej Intagorn for his help in retrieving data from Digg and Aram Galstyan for useful discussions. This work is supported in part by National Science Foundation under award 0915678.


\end{document}